\documentstyle[aps,multicol,epsfig]{revtex}

\newcommand{\be}{\begin{equation}}
\newcommand{\ee}{\end{equation}}

\begin{document}
 
\draft

\title{Noise--Induced Phase Separation: Mean--Field Results}
\author{M. Iba\~nes$^{1}$, J. Garc\'{\i}a-Ojalvo$^{2,3}$, R. Toral$^4$ and 
J.M. 
Sancho$^{1}$}
 
\address{$^{1}$
Departament d'Estructura i Constituents de la Mat\`eria, 
Universitat de Barcelona\\ Diagonal 647, E--08028 Barcelona, Spain\\
$^2$ Departament de F\'{\i}sica i Enginyeria Nuclear, Universitat
Polit\`ecnica de Catalunya\\
Colom 11, E--08222 Terrassa, Spain\\
$^3$ Institut f\"ur Physik, Humboldt Universit\"at zu Berlin,
Invalidenstr. 110, D--10115 Berlin, Germany\\
$^4$ Instituto Mediterr\'aneo de Estudios Avanzados (IMEDEA, CSIC-UIB) 
and Departament de F\'{\i}sica,
Universitat de les Illes Balears,
E--07071 Palma de Mallorca, Spain\\
 }
 
\date{\today }
\maketitle

\begin{abstract}
 
We present a study of a phase--separation process induced by the presence 
of spatially--correlated multiplicative noise. We develop a mean--field 
approach suitable for
conserved--order--parameter systems and use it to obtain the
phase diagram of the model. Mean--field results are compared with numerical
simulations of the complete model in two dimensions. Additionally, a comparison
between the noise--driven dynamics of conserved and nonconserved systems
is made at the level of the mean--field approximation.
 
\vspace{8pt}
PACS: 05.40.-a, 64.60.-i
\vspace{8pt}
 
\end{abstract}

\begin{multicols}{2}

\section{Introduction}

Many theoretical and experimental observations confirm nowadays the existence
of noise--induced order. Phenomena such as noise--induced transitions
\cite{horsthemke84}, stochastic resonance \cite{gamma98} and noise--induced
transport \cite{hanggi96} are examples of the ordering features of
fluctuations in purely temporal dynamical systems. Additionally, recent
years have witnessed an increasing interest on noise--induced phenomena
in spatially--extended systems (see \cite{nises99} for a recent review).
Some of the topics studied in this respect include noise--induced
patterns \cite{ojalvo93,parr96}, noise--induced phase transitions
\cite{VPT,jordi96,man97}, spatiotemporal stochastic resonance \cite{SSR1,SSR2},
noise--induced fronts \cite{santos99}, noise--supported traveling
structures in excitable media \cite{exc} and noise
sustained convective structures \cite{dei89,SCSW}. 
We are concerned in this paper
with the phenomenon of noise--induced phase separation, recently observed
in systems with conserved dynamics \cite{jordi98}.

Several analytical methods have been used so far \cite{nises99} to
examine the above--mentioned spatiotemporal problems. By way
of example, the stability of a homogeneous state with respect to
small perturbations of arbitrary wavenumber can be analyzed
in a linear approximation. Such a linear stability analysis shows
that pattern--forming transitions are nontrivially affected by
multiplicative noise \cite{ojalvo93,becker94}. From a more fundamental
point of view, systems exhibiting phase transitions in a
statistical--mechanics sense can be investigated by means of the
dynamic renormalization group \cite{DRG1,DRG2}, which shows that under
certain conditions a new genuine nonequilibrium universality class
arises due to the presence of multiplicative noise \cite{MNUC1,MNUC2}. 
A third fruitful approach is based on the well--known mean--field
approximation, widely used in the context of equilibrium statistical
mechanics, and that has been recently extended to nonequilibrium
systems under the influence of external noise \cite{VPT,broeck94}.
In this context, the approximation is introduced by assuming that
the interaction between a certain spatial point and its neighbors
occurs through a mean value of the field, which corresponds to its
statistical average at the given point. This approach
has led to the prediction of noise--induced ordering and disordering
phase transitions (NIOTs and NIDTs), which has been successfully
verified (at least qualitatively) by numerical simulations in different
models \cite{VPT,jordi96,MNUC2,zaikin}. The advantages of this procedure
as compared to, e.g., linear--stability approaches, lie on its ability
to describe the system arbitrarily far from the transition point
and to take into account the influence of spatial coupling strength,
that arises naturally in discretized systems. In this way, the
mean--field analysis can successfully explain the existence of
successive NIOTs and NIDTs (also called reentrant transitions in
this context) as a single control parameter is varied.

The aim of this paper is to perform a somewhat detailed study,
using the mean--field approximation technique, of the phenomenon of
noise--induced phase separation. This phenomenon has been recently
predicted by a linear stability approach and confirmed by numerical
simulations \cite{nises99,jordi98}. It arises in spatiotemporal systems
whose dynamics is conserved, in the sense that the spatial average of the
field does not vary with time, but depends only on the initial conditions
of the system. Due to this fact, a standard mean--field approach cannot
be applied in this case, because no change in the mean field will be
observed as a given control parameter is varied (and hence no phase
transition can be found in this way). Therefore, an extension of the
procedure is needed in order to handle this situation. The
present work is devoted to developing such an extension, and applying
the results to the particular case of noise--induced phase separation
mentioned above. The outline of the rest of the paper is the following:
Section II introduces the general system that will be investigated,
along with the particular model to which the obtained results will
be applied. A comparison between conserved and nonconserved dynamics
is also briefly sketched. Section III reviews the mean--field procedure
for nonconserved systems, and extends it to include the effect of
spatial correlation of the external noise. Section IV introduces
the generalized mean--field approach for conserved systems.
Section V discusses the
limit of strong spatial coupling of the procedure, and compares
the corresponding results with those coming from linear stability
analysis. Throughout all these Sections, a comparison with respective
numerical simulations of the complete model is made. Finally, some
conclusions are stated in Section VI.

\section{Conserved and nonconserved dynamical models}

The spatiotemporal dynamics of a nonequilibrium system in the
presence of both internal and external noise can be described by
the following Langevin equation \cite{Ito} for the time evolution
of the $d$--dimensional scalar field $\phi(\vec x,t)$:
\begin{equation}
\frac{\partial \phi(\vec x,t)}{\partial t} = -\Gamma\left[\frac{\delta F}
{\delta \phi} + g(\phi)\,\xi(\vec x,t)\right] + \eta(\vec x,t)\,,
\label{eq:mf-spdem}
\end{equation}
where the additive noise $\eta(\vec x,t)$ is Gaussian and white, with
zero mean and correlation
\begin{equation}     
\langle \eta(\vec x,t)\,\eta(\vec x',t')\rangle=2\,\varepsilon\,\Gamma\,
\delta(\vec x-\vec x')\,\delta(t-t')\,.
\label{eq:mf-wnc2}
\end{equation}
The intensity of the noise is measured by the parameter $\varepsilon$.
The existence of the factor $\Gamma$ in correlation (\ref{eq:mf-wnc2})
is a sign of the internal character of this noise, in whose only presence 
($g=0$)
the system can exhibit {\em equilibrium} properties. The multiplicative
noise term $\xi(\vec x,t)$, on the other hand, is external and brings
the system {\em out of equilibrium}. It may arise, for instance, from a 
fluctuating control parameter. It is also Gaussian with zero mean, but its
correlation will be assumed in principle to have a nontrivial
structure in space: 
\begin{equation}
\langle \xi(\vec{x},t)\,\xi(\vec{x'},t')\rangle=2\,\sigma^2\,
c(|\vec{x}-\vec{x'}|)\,\delta(t-t')\,,
\label{eq:ncmcol}
\end{equation}
where $c(|\vec{x}-\vec{x'}|)$ is the spatial correlation 
function of the external noise and  $\sigma^2$ is its 
intensity.

Different and physically motivated choices for $\Gamma$ will lead to 
a variety of dynamical and 
steady state phenomenologies.
The particular case of $\Gamma=-\nabla^2$ (called {\em model B} in the 
literature of critical phenomena) is appropriate to describe a system in 
which  the global quantity $\int \phi(\vec x,t)d^d\vec x$ is conserved in time. 
Physical realizations of this system include the case of phase separation 
in binary alloys. In this case, an 
initial mixture of the two components may undergo, for some values of the 
control parameters, a separation process which,
depending on the initial relative concentrations of each
component,  takes the form of spinodal decomposition or nucleation
\cite{gunton83}. In this paper we will be mainly concerned with the 
conserved case, although 
 a comparison will also be made with the
corresponding nonconserved case, defined by $\Gamma=1$ (known as  
{\em model A} ).  

Even though the theoretical approach that will be presented here is quite 
general, our results
will be applied, for the sake of clarity, to the particular  
Ginzburg--Landau form of the free energy $F$,
\begin{equation}
\label{eq:fe}
F=\int dx\left[ V(\phi) +\frac{D}{4d}
|\vec \nabla\phi|^2\right]\,,
\end{equation}
where the local potential $V(\phi)$ is
\begin{equation}
\label{eq:pot}
V(\phi)=-\frac{a}{2}\,\phi^2+\frac{1}{4}\,\phi^4\,.
\end{equation}
In the absence of noise sources, the behavior of this potential
is the following: for $a\le 0$ the homogeneous trivial solution $\phi=0$ is
stable, whereas for $a>0$ that solution becomes unstable.
This instability gives rise either to a phase transition
towards an ordered (ferromagnetic) phase in the nonconserved case, or to
a phase separation process in the conserved case. 

The external noise will be taken to be coupled to the field according to
\begin{equation}
g(\phi)=\phi\,,
\label{eq:fandg}
\end{equation}
which corresponds to allow 
the control parameter $a$ in (\ref{eq:pot}) to fluctuate in space and time.
We will use the following Gaussian spatial correlation function
\begin{equation}
\label{eq:ccol}
c\left(|\vec x-\vec x'|\right)=\frac{1}
{(\lambda\sqrt{2\pi})^d}\,
\exp\left(-\frac{|\vec x-\vec x'|^2}{2 \lambda^2}\right)\,,
\end{equation}
whose width $\lambda$ characterizes the correlation length of the noise.
The normalization is such that in the limit $\lambda \to 0$ this 
correlation 
goes to a delta function and $\xi(\vec x,t)$ becomes a spatial white noise 
with intensity $\sigma^2$.

\section{Mean--field approach for nonconserved dynamics}

We now review the main points of the mean--field approach in its application
to nonconserved order--parameter systems (model A), in order to clarify
the extension to conserved dynamics that will be presented in the next
Section. We begin by discretizing the field equation (\ref{eq:mf-spdem})
with $\Gamma=1$ in a regular 
$d$--dimensional lattice of mesh size $\Delta x = 1$, and lattice
points ${\vec x_1,\dots,\vec x_N}$
\begin{equation}
\frac{d \phi_i}{ d t} = f(\phi_i) + \!\frac{D}{2d} \sum_{j}
\,\tilde D_{ij}\,\phi_j +
 \eta_i(t)\, + g(\phi_i)\,\xi_i(t)\,,
\label{eq:spdedis}
\end{equation}
where $\phi_i \equiv \phi(\vec x_i)$, $f(\phi_i)=-V'(\phi_i)$, 
and only one index is used to label the cells,
independently of the dimension of the lattice.
$\tilde D_{ij}$ accounts for the discretized Laplacian operator
\begin{equation}
\label{eq:laplaop}
\nabla^2 \,\to \,\sum_j\,\tilde D_{ij}=\sum_j\left( 
\delta_{nn(i),j}-\,2d\,\delta_{i,j}\right)\,,
\end{equation}
where $nn(i)$ represents the set of all the sites which are 
nearest neighbors of site $i$.

The discrete noises $\eta_i(t)$ and $\xi_i(t)$
are still Gaussian with zero mean and their correlations are given by
\begin{equation}
\langle \eta_i(t)\,\eta_j(t')\rangle=2\,\varepsilon\,
\delta_{i,j} \,\delta(t-t')
\label{eq:wnc2dis}
\end{equation}
and
\begin{equation}
\langle \xi_i(t)\,\xi_j(t')\rangle=2\,\sigma^2\,
c_{|i - j|}\,\delta(t-t')\,,
\label{eq:ncmcold}
\end{equation}   
where $c_{|i - j|}$ is a convenient discretization of the function
$c(|\vec x-\vec x'|)$  and specific values such as $c_0,~ c_1$
have to be computed numerically \cite{nises99,jordi98} when needed.
For the white noise case ($\lambda=0$) and with the mesh size
chosen $\Delta x=1$ it is $c_0=1,\, c_1=0$, whereas for large $\lambda$,
$c_0$ scales roughly as $c_0 \propto \lambda^{-d}$. 

The  corresponding Fokker--Planck equation, in the Stratonovich 
interpretation, 
for the multivariate probability density $P(\phi_1,\phi_2,\ldots,t)\equiv
P(\{\phi\},t)$ is \cite{nises99}
\begin{eqnarray}
\label{eq:fpcol}
\frac{\partial P}{\partial t} = -\sum_i\frac{\partial}
{\partial \phi_i} \left[ f(\phi_i) + \frac{D}{2d} \sum_{j\in nn(i)}
(\phi_j-\phi_i)\,  -\right.
\nonumber\\
\left.
-\,\varepsilon\,\frac{\partial}{\partial\phi_i} 
-\sigma^2\, g(\phi_i)\,
\sum_j\,c_{|i-j|}\,\frac{\partial}{\partial\phi_j} \, g(\phi_j) 
\right] \,P\,. 
\end{eqnarray}

In order to get the  evolution equation for the single--site probability 
distribution $P(\phi_i,t)$, defined as
\begin{equation}
\label{eq:mf-sppd}
P(\phi_i,t)=\int\left[\prod_{k\neq i}d\phi_k\right] P(\{\phi\},t)\,,  
\end{equation}
we integrate Eq.~(\ref{eq:fpcol}) over all the variables except $\phi_i$.
Vanishing of the probability for the field going to $\pm \infty$ leads to 
\begin{equation}
\label{eq:pista}  
\int\left[\prod_{k\neq i}d\phi_k \right]
\frac{\partial \,}{\partial\phi_j}\,\left[g(\phi_j)\,P(\{\phi\},t)\right]=0
\qquad j\neq i
\end{equation}
and using the standard definition of the conditional probability, one 
gets  
\begin{eqnarray}
\sum_{j\in nn(i)}\int \left[\prod_{k\neq i}\,d\phi_k \right]\,\phi_j 
\,P(\{\phi\},t)\equiv
\label{eq:mf-cond-av}
\nonumber\\
\left[\sum_{j\in nn(i)} \int d\phi_j \,\phi_j
\,P(\phi_j|\phi_i,t) \right] P(\phi_i,t) \equiv 
\nonumber\\
2d\,\langle \phi(t) 
\rangle_{\phi_i}\,P(\phi_i,t)
\end{eqnarray}
which defines $\langle\phi(t)\rangle_{\phi_i}$ as a nearest-neighbor
conditional average.
Thus we finally find that the one point steady probability distribution 
follows the simpler but still exact equation,
\begin{eqnarray}
\label{eq:statcol}
\frac{\partial P(\phi_i,t)}{\partial t} = - \frac{\partial 
\,}{\partial \phi_i} \left[f(\phi_i) + 
D\left(\langle\phi(t)\rangle_{\phi_i}-\phi_i\right) - \right. 
\nonumber\\
\left.
-\,\varepsilon\,\frac{\partial \,}{\partial\phi_i}
-\,\sigma^2\,c_0\,g(\phi_i)\,
\frac{\partial \,}{\partial \phi_i}\,g(\phi_i) \,\right] P(\phi_i,t)\,.
\end{eqnarray}

The mean--field approximation consists in assuming that the 
conditional average in the last equation is replaced by \cite{text}  
\begin{equation}
\label{eq:meanf}
\langle\phi(t)\rangle_{\phi_i} =
\langle\phi_i(t)\rangle\,,
\end{equation}
which is equivalent to doing directly the following assumption at the level
of the Langevin Eq.~(\ref{eq:spdedis}),
\begin{equation}  
\label{eq:meanf2}
\frac{1}{2d}\sum_j\,\tilde D_{ij}\,\phi_j(t)=\langle\phi_i(t)\rangle -
\phi_i(t)\,.
\end{equation}
Using this approximation, and imposing the condition of stationary probability 
distribution with no flux, we
get that the single--site steady distribution satisfies,
\begin{eqnarray}
\label{eq:statcol1}
\left[f(\phi) + D\left(\langle\phi\rangle_{\rm st}-\phi\right)
-\,\varepsilon\,\frac{\partial \,}{\partial\phi}\, - \right.
\nonumber\\
\left.
-\,\sigma^2\,c_0\,g(\phi)\,
\frac{\partial \,}{\partial \phi}\,g(\phi) \right] P_{\rm st}(\phi) = 0\,,
\end{eqnarray}  
where subscript $i$ has been dropped for simplicity.

The solution of the previous equation can be easily written
down:
\begin{eqnarray}
\label{eq:pcol}   
P_{\rm st}(\phi,\langle\phi\rangle_{\rm st}) = N\exp \int 
d\phi'\,\frac{1}{\sigma^2\,c_0\,g^2(\phi')+\varepsilon}\times
\nonumber\\
\nonumber\\
\left[f(\phi')+D\,(\langle\phi\rangle_{\rm st}-\phi')-
\sigma^2\,c_0\,g(\phi')\,g'(\phi')\right]\,, 
\end{eqnarray}
where $N$ is an appropriate normalization constant. The above solution
is only formal, because $\langle\phi\rangle_{\rm st}$ depends on the 
probability distribution
itself. 
However, 
both $P_{\rm st}$ and $\langle\phi\rangle_{\rm st}$
can be determined by means of the following self--consistency
relation, which is a 
signature of the mean--field approach,
\begin{equation}
\label{eq:mf-scr}
\langle \phi \rangle_{\rm st} =\int_{-\infty}^\infty \phi\, P_{\rm st}(\phi, 
\langle \phi \rangle_{\rm st})\,d\phi\,. 
\end{equation}

We now apply our results to the particular model defined by
(\ref{eq:fe})--(\ref{eq:ccol}).
The solution of the self--consistency
equation (\ref{eq:mf-scr}) in this case is plotted in Fig.~\ref{fig1}
versus the control parameter $a$
for three different values of the noise correlation length $\lambda$.
Note that the existence of multiplicative noise shifts the critical
point towards negative values of $a$, which indicates the ordering
character of the external noise. This noise--induced phase transition
has been substantially studied in the past in the case of white external
noise \cite{jordi96,becker94,broeck94}.
Figure \ref{fig1} also displays the results obtained by numerical simulations
of the complete model (\ref{eq:spdedis}) for a two--dimensional square lattice  and using the mean--field results as the initial conditions.
It can be seen
that mean--field results give the correct qualitative behavior of the system,
and are also quantitatively right far from the critical point. The agreement
between mean--field predictions and simulations close to the critical point
improves when the correlation
length $\lambda$ increases. 

Notice that, in this mean--field approximation,
the only effect of the finite correlation length $\lambda$ of the noise shows up
in the value $c_0< 1$ (see Eq.~(\ref{eq:pcol})) of the correlation function at zero distance. As mentioned
before, this value decreases with increasing $\lambda$.
In other words, 
for nonconserved dynamics the disordering effect of
the spatial correlation of multiplicative noise in the mean-field approximation
arises only through a decrease
of the effective noise intensity.

\section{Mean--field approach for conserved dynamics}

The mean--field approach discussed above cannot be straightforwardly 
extended to deal with conserved--order--parameters systems, because
in these cases the mean field $\langle\phi\rangle$ is constant in
time, depending only on the initial conditions and not on the control
parameter $a$. We now introduce a generalized mean--field approximation
that overcomes such a restriction. The main ideas underlying this
extension will be first presented in the deterministic model B.

\subsection{Deterministic dynamics}

In the absence of all noise sources, model B takes the form
\begin{equation}
\label{eq:b}
\frac{\partial \phi(\vec{x},t)}{\partial t}= \nabla^2\frac{\delta F}
{\delta \phi}\,.
\end{equation}
This model evolves in time under the following restriction
\begin{equation}
\label{eq:conserv}
\frac{1}{V}
\int d\vec{x}\,\phi(\vec{x},t)=\phi_0 \,, 
\end{equation}
being $\phi_0$ fixed by the initial conditions.
The phenomenology of this model is well known \cite{gunton83}: there is a 
transition
point $a_T(\phi_0)$, such that
for $a<a_T(\phi_0)$ the homogeneous state $\phi=\phi_0$ is stable, whereas for
$a>a_T(\phi_0)$ the system separates in two bulk phases, $\phi_1$ and $\phi_2$,
fulfilling that the spatial average of $\phi$ is also equal to $\phi_0$.
The transition from an homogeneous state to a two--phase state is 
critical ({\em i.e.}, of second order) for $\phi_0=0$, so that
$a_T(\phi_0=0)\equiv a_c$.

In order to determine both $a_T(\phi_0)$ and $a_c$, we look for
the steady--state solutions of Eq.~(\ref{eq:b}). These solutions
fulfill the Laplace equation $\nabla^2\frac{\delta F}{\delta \phi} = 
0$. The analytical and bounded solution is 
$\frac{\delta F}{\delta \phi} = h$, being $h$ a constant. 
Therefore, the steady
states of model B can be interpreted as the minima of an effective potential
$F_{\rm eff}=F-h\int d\vec{x}\phi$, 
and coincide with the steady states of
model A with an external control field $h$. 
Following reference \cite{miguela} we call $h$ the
{\em constant effective field} of the system. For equilibrium systems  
 $h$ is nothing but the chemical   
potential. Moreover, $h$ is not an arbitrary
constant, and it has to be determined by imposing the conservation law, 
Eq.~(\ref{eq:conserv}).
Substitution of the Ginzburg-Landau form in the discretized version of 
$h = \frac{\delta F}{\delta \phi}$, leads to
\begin{equation}
\label{eq:modbdet}
h =  -a\phi_i+\phi_i^3 - 
\frac{D}{2d}\sum_j\,\tilde D_{ij}\,\phi_j \,.
\end{equation}

We now need to consider separately the subthreshold and superthreshold
situations:
\begin{itemize}
\item In the subthreshold (homogeneous) case the condition 
$\phi_i=\phi_0, \forall i$,
has to be verified, and therefore Eq.~(\ref{eq:modbdet}) reduces to
\begin{equation}
\label{eq:h}
h= -a\phi_0+\phi_0^3\,.
\end{equation}
Hence the value of $h$ does depend on the initial condition in the
subthreshold situation.
\item Above the transition point (not yet determined), the steady state of the 
system is not globally homogeneous, since the field separates in two bulk phases
with values $\phi_1$ and $\phi_2$, respectively.
The fraction $x$ of system in phase $\phi_1$ is given by
the lever rule: $x \phi_1 +(1-x)\phi_2=\phi_0$. 
For a general free 
energy, a Maxwell-type construction would give us the value of $h$. In the 
case of a locally
symmetric free--energy (such as the one defined in Eqs.~(\ref{eq:fe})) and
(\ref{eq:pot}), a simpler argument can be used: 
each phase has to satisfy 
Eq.~(\ref{eq:modbdet}) with $\phi_i$ equal to the 
field value
of the corresponding phase, either $\phi_1$ or $\phi_2$, and, since by 
the symmetry of the free energy, these two quantities verify 
$\phi_1=-\phi_2$, 
$h$ must be zero. Consequently we get  
\begin{equation}
\label{eq:h0}   
\phi_{1,2} = \pm \sqrt{a} \,,
\end{equation}
which are the solutions of the deterministic model A for a value of the
external control field $h=0$. 
\end{itemize}
Just at the transition point, there is a unique phase $\phi=\phi_0$ 
and $h$ is identically zero. Thus the transition line (also called in this
context coexistence line) is given by 
\begin{equation}
a_T( \phi_0) = \phi_0^2\,.
\end{equation}
We also note that for $\phi_0=0$, the critical
point is obtained: $a_T=a_c=0$.

We will now show that the concept of the constant effective field can
be used to generalize the mean--field approximation to conserved
systems {\em with noise}.

\subsection{Noise--induced phase separation}

We now add stochastic sources to model (\ref{eq:b}), in the form of
both an internal additive noise and an external multiplicative one.
The resulting model is represented by
Eqs.~(\ref{eq:mf-spdem})--(\ref{eq:ncmcol}) with $\Gamma=-\nabla^2$.
The discretized version of this model is
\begin{eqnarray}
\frac{d \phi_i}{d t}&=&
-\sum_j \tilde D_{ij}\left(f(\phi_j) +
\frac{D}{2d}
\sum_k \tilde D_{jk}\,\phi_k  + \, g(\phi_j)\,\xi_j \right)+
\nonumber
\\
\label{eq:mf-bspdem}
& &+\, \eta_i(t)\,,
\end{eqnarray}
with $f(\phi_j)=-V'(\phi_j)$, as before. 
The correlation of the additive noise is now
\begin{equation}
\langle \eta_i(t)\,\eta_j(t')\rangle=-2\,\varepsilon\,\tilde D_{ij}\,
\delta(t-t')\,,
\label{eq:ruidocd}
\end{equation} 
and that of the multiplicative noise was already introduced in Eq.~(\ref
{eq:ncmcold}).
The corresponding Fokker-Planck equation, in the Stratonovich interpretation,
 for the multivariate
probability density $P(\{\phi\},t)$ 
is in this case
\begin{eqnarray}
\frac{\partial P}{\partial t} =\sum_{i,j}\frac{\partial}
{\partial \phi_i}\tilde D_{ij}  \left(f(\phi_j) + \frac{D}{2d}
\sum_{k}\tilde D_{jk} \,\phi_k\, - \right.
\nonumber\\
\left.
- \,\varepsilon\,\frac{\partial \,}{\partial \phi_j}
+\sigma^2\,  
g(\phi_j)\,\sum_{r,s}\,\frac{\partial \,}{\partial \phi_s}\,
\tilde D_{sr}\,c_{|j-r|}\, 
g(\phi_r)\right)P\,.
\label{eq:fpcolb}
\end{eqnarray}
As done in the nonconserved case, we now integrate Eq.~(\ref{eq:fpcolb}) 
over all the variables except $\phi_i$, in order to get the evolution equation
of the single--site probability distribution $P(\phi_i,t)$ (see 
Eq.~(\ref{eq:mf-sppd}))
\begin{equation}
\label{eq:fpcolb1}
\frac{\partial P(\phi_i,t)}{\partial t} =\frac{\partial}{\partial \phi_i}
\sum_j \tilde D_{ij}\langle M_j \rangle_{\phi_i}\,P(\phi_i,t) \,,
\end{equation}
where 
\begin{eqnarray}
M_j=f(\phi_j) + \frac{D}{2d}\sum_{k}\tilde D_{jk} \,\phi_k\,
- \,\varepsilon\,\frac{\partial \,}{\partial \phi_j}+
\nonumber\\
+\,\sigma^2\,g(\phi_j)\,\sum_{r,s}\,\frac{\partial \,}{\partial \phi_s}\,
\tilde D_{sr}\,c_{|j-r|}\,g(\phi_r) \,.
\label{eq:M}
\end{eqnarray}
If we impose the condition of stationarity probability distribution with
no flux, $\langle M_j \rangle_{\phi_i}$ must satisfy 
\begin{equation}
\sum_j \tilde D_{ij}\langle M_j \rangle_{\phi_i}\,P_{\rm st}(\phi_i)=0\,.
\label{eq:laplace}
\end{equation}
In the deterministic conserved case it has been shown that the solution of 
this equation is the constant effective field,
$\langle M_j \rangle_{\phi_i}=-h$.
We can now take $j=i$ and perform the conditional average of $M_i$.
If we  
consider the expression analogous to Eq.~(\ref{eq:pista}) for the 
multiplicative noise term
and make the standard mean field approximation
(\ref{eq:meanf}) 
we arrive at
\begin{eqnarray}
\label{eq:pint}
-h\,P_{\rm st}(\phi)=\left(f(\phi) + D\,\left(\langle \phi
\rangle_{\rm st} -\phi\right) -\,\varepsilon\,\frac{\partial \,}{\partial\phi}
\,+ \right.\qquad
\nonumber\\
\left.
+2d\,\sigma^2g(\phi)\left[c_1\, g(\langle \phi\rangle_{\rm st})
\,\frac{\partial\,}{\partial \phi}
-c_0
\frac{\partial\,}{\partial \phi}g(\phi)\right]\right)P_{\rm st}(\phi)\,,
\end{eqnarray}
where subscripts have been dropped again for simplicity. In the derivation
of this equation a generalization of the mean--field approximation
for the nearest--neighbor conditional average of function $g(\phi)$ has been 
applied, namely:
\begin{equation}
\langle g(\phi) \rangle_{\phi_i} =  g \left(\langle \phi_i\rangle \right)\,.  
\end{equation}
In principle this is an uncontrolled approximation whose validity needs to be
assessed by the numerical simulations which will be presented in what follows.

The solution of Eq.(\ref{eq:pint}) yields the following stationary 
probability distribution 
\begin{eqnarray}
\label{eq:pcolbg}
P_{\rm st}(\phi,\langle\phi\rangle_{\rm st},h)=
\qquad\qquad\qquad\qquad
\qquad\qquad\qquad
\nonumber\\
\nonumber\\
=N \exp\int \frac{d\phi'}{2d\sigma^2\,g(\phi')\left(
c_0\,g(\phi')-c_1g(\langle \phi\rangle_{\rm st})\right)+\,\varepsilon}\times
\nonumber\\
\nonumber\\
\left[f(\phi')+D\left(\langle\phi\rangle_{\rm st}
-\phi'\right)-2d\sigma^2c_0g(\phi')g'(\phi')+ h\right]\,,
\end{eqnarray}
where $h$ and $\langle\phi\rangle_{\rm st}$ are parameters to be determined
self-consistently.

We now particularize the result obtained above to the Ginzburg--Landau
model defined by Eqs.~(\ref{eq:fe})--(\ref{eq:ccol}). In this case,
the stationary single--site probability distribution is
\begin{eqnarray}
\label{eq:pcolb}
P_{\rm st}(\phi,\langle\phi\rangle_{\rm st},h)=N\times
\qquad\qquad\qquad\qquad\qquad
\qquad
\nonumber\\
\nonumber\\
\exp\int\frac{\left(a - D -2d\sigma^2c_0 \right)\phi'-\phi'^3 + 
D\langle\phi\rangle_{\rm st} + h}{2d\,\sigma^2\left( 
c_0\phi'^2
-c_1\langle\phi\rangle_{\rm st}\,\phi'\right)+\varepsilon}\,d\phi'\,.
\end{eqnarray}

We now need to determine the unknown constants $h$ and $\langle
\phi\rangle_{\rm st}$. Similarly to the deterministic case studied above,
we consider separately the subthreshold and superthreshold 
situations. We recall at this point that the mean--field approach
presented above is {\em local}, and leads to an expression for
the probability distribution of the field at a given site of
the lattice as a function of $h$ and of the
mean field $\langle\phi\rangle_{\rm st}$ in the neighborhood of the
given cell. In the homogeneous case ($a<a_T$), this mean field
is the same everywhere, and it is equal to $\phi_0$. Hence
only $h$ is left to be evaluated, what can be done by means
of the generalized self--consistency relation
\begin{equation}
\label{eq:mf-scrb}
\langle \phi \rangle_{\rm st} = \int_{-\infty}^\infty \phi\,
P_{\rm st}(\phi, \langle \phi \rangle_{\rm st},h)\,d\phi \,.
\end{equation}
 with $\langle \phi \rangle_{\rm st} = \phi_0$.  

For $a>a_T$ the system has two phases, and thus there are two different
local mean values, corresponding to each of the bulk phases
($\langle\phi\rangle_1$ and $\langle\phi\rangle_2$).
Because of the symmetry of $F$ and $g$, these values must satisfy
$\langle\phi\rangle_1=-\langle\phi\rangle_2$. Therefore, since $h$
needs to be the same for the two phases, and given the form of the
numerator in (\ref{eq:pcolb}), $h$ must be zero in this ordered state.
Hence only the values of the local (symmetrical) mean fields
$\langle\phi\rangle_1$ and $\langle\phi\rangle_2$ need to be
determined for $a>a_T$. This can be done as in the case of model
A, solving the self--consistency relation (\ref{eq:mf-scr}) using the 
steady probability given in (\ref{eq:pcolb}) with $h=0$. For nonsymmetric 
functions $F(\phi)$ and $g(\phi)$ a possible extension of the Maxwell rule
is to choose $h$ in such a way that the two solutions of Eq.(\ref{eq:mf-scrb})
have the same value for the stationary probability 
$P_{\rm st}(\langle \phi \rangle_{\rm st},\langle \phi \rangle_{\rm st},h)$.

The bifurcation diagram resulting from the application
of the self--consistency relations  
is plotted in Fig.~\ref{fig2} for three different values of
the multiplicative--noise correlation length. 
Numerical simulation
results of the complete model (\ref{eq:mf-bspdem})
are also shown. Mean field results have been used as the 
initial conditions of the numerical simulations letting each phase evolve 
until its stationary value. 
The effects of the intensity and correlation
length of the multiplicative noise are qualitatively the same as in model
A: the noise--induced shift of the transition point, in the direction of
enhancing order in the system, increases with noise intensity and decreases
with correlation length. 

Figure~\ref{fig3} shows the values of the constant effective field 
$h$ obtained numerically by imposing the self--consistency relation
(\ref{eq:mf-scrb}) until $h$ vanishes, for a nonzero initial concentration
$\phi_0=0.2$. The corresponding value of the
control parameter $a$ at which $h$ first becomes zero is the transition point
$a_T$. Results have been plotted for the deterministic case 
(which can be calculated analytically, see Eq.~(\ref{eq:h})), 
the case with just additive
noise, and three cases with also multiplicative noise for different
correlation lengths, corresponding to the situations shown
in Fig.~\ref{fig2}. The noise--induced shift of the transition point
and the influence of the noise correlation length as well as the
disordering role of the additive noise (reflected
in the shift of the transition point towards the right when only additive noise
is considered -- dotted line) can be clearly seen.

A comment on the comparative influence of multiplicative noise on conserved and 
nonconserved
dynamics is worth to be made at this point. We note that, in the
ordered state ($h=0$), the single--site probability distribution
of the conserved model (\ref{eq:pcolbg}) in the presence of {\em white}
multiplicative noise ($c_0=1$ and $c_1=0$) reduces to 
\begin{eqnarray}
\label{eq:pcolbgw}
P_{\rm st}(\phi,\langle\phi\rangle_{\rm st})=N
\exp\int d\phi'\frac{1}{2d\sigma^2\,g^2(\phi')+\varepsilon}\times
\qquad
\nonumber\\
\nonumber\\
\left[f(\phi')+D\left(\langle\phi\rangle_{\rm st}-\phi'
\right)-2d\sigma^2g(\phi')g'(\phi')\right] \,,
\end{eqnarray}
which should be compared with the corresponding expression (\ref{eq:pcol})
for the nonconserved case with $c_0=1$. One can easily
see that multiplicative noise
has a stronger effect on the conserved model than on the nonconserved one, 
since in the former
case the noise intensity is multiplied by a factor $2d$.
In the particular case in which the two
noise intensities of the nonconserved ($A$) and conserved ($B$) models are related by
\begin{equation}
\sigma^2_{A}=2d\,\sigma^2_B\,,
\label{eq:sigma}
\end{equation} 
the two models are equivalent above the transition point. However, this
equivalence disappears in the case of {\em colored} multiplicative noise,
because of the term $2d\,\sigma^2\,c_1g(\phi')g( \langle \phi\rangle_{\rm st})$
appearing in
Eq.~(\ref{eq:pcolbg}). This different 
dependence indicates that spatial
correlation of the noise is more relevant for the conserved model 
than for the nonconserved one,
where correlation length of the noise produces only a shift of the
transition point \cite{jordi98}. A comparison between the results of 
model A and B is shown in Fig.~\ref{fig4}, for both $\lambda=0$ and
$\lambda \neq 0$. Noise intensities have been chosen here to verify 
expression (\ref{eq:sigma}), so that in the white--noise case, mean--field 
results coincide for the two models. Mean-field results are in better
agreement with simulations in the case of model B. 

Finally, we now address the issue of whether a reentrant noise--induced
phase transition towards disorder arises in the conserved model B. Previous works have
shown the existence of such a transition for nonconserved models
\cite{VPT,jordi96}. This means that for fixed values of $a$ and $D$, when
increasing the multiplicative noise intensity, the system goes first through
a phase transition from disorder to order 
(NIOT) and then, for higher values of the noise, it experiments another
transition back to disorder (NIDT).
These two transitions can only be found when increasing $\sigma^2$, instead
of $a$ or $D$. 

Mean--field theory predicts the existence of reentrant
transitions also for model B, as shown in Fig.~\ref{fig5}. This figure
shows the behavior of the mean field $\langle\phi\rangle_{\rm st}$ versus
multiplicative noise intensity with two different correlation lengths 
($\lambda=0$ and $\lambda=0.5$) for model A and B. Again, the noise
intensities for the two models have been chosen to verify (\ref{eq:sigma}),
so that the $\lambda=0$ result is identical in the two cases. However,
the effect of the correlation length is different for the two models:
whereas for model A $\lambda$ retards both the NIOT and the NIDT,
for model B the NIOT is retarded, but the NIDT is advanced. This is
an indication of the nontrivial influence of the noise correlation
length in the conserved case.

\section{Strong coupling limit}

In the limit of strong coupling, $D\to\infty$, the predictions of
mean--field theory can be evaluated analytically and should agree with
the results given by a standard linear stability analysis of the model.
In order to verify this agreement, we will now compute this limit
for the mean--field results obtained so far, for both models A and B.

\subsection{Model A}

In the mean--field approximation and in the limit $D\to\infty$,
the stationary probability distribution 
$P_{\rm st}(\phi,\langle\phi\rangle_{\rm st})$
(\ref{eq:pcol}) becomes
\begin{equation}
\label{eq:pst}   
P_{\rm st}(\phi,\langle\phi\rangle_{\rm st})=\delta(\phi-
\langle\phi\rangle_{\rm st})\,,
\end{equation}
as can be easily seen by means of a steepest--descent calculation.
This expression verifies trivially the self-consistency relation
(\ref{eq:mf-scr}), which can thus no longer be used to determine
$\langle\phi\rangle_{\rm st}$. In order to do that, we now integrate
Eq.~(\ref{eq:statcol1}) with respect to $\phi$, and obtain
\begin{equation} 
\label{eq:mf-spdemmediadanovst}
\langle f(\phi) \rangle_{\rm st} +
\sigma^2\,c_0\, 
\langle g'(\phi)\, 
g(\phi)\rangle_{\rm st}=0\,.
\end{equation}
For $D\to\infty$, these averages are evaluated trivially using
expression (\ref{eq:pst}), and Eq.~(\ref{eq:mf-spdemmediadanovst}) becomes
\begin{equation}
\label{eq:conmedio}
f(\langle \phi \rangle_{\rm st} ) + \sigma^2\,c_0\, g'(\langle \phi 
\rangle_{\rm st})
\,g(\langle \phi \rangle_{\rm st} )=0\,,
\end{equation}
from which $\langle \phi \rangle_{\rm st}$ can be found.
For model A and in the case defined by Eqs.~(\ref{eq:fe})--(\ref{eq:ccol}),
the solutions of this equation are either
\begin{equation}
\label{eq:f1} 
\langle \phi \rangle_{\rm st}=0
\end{equation}   
or
\begin{equation}
\label{eq:f2}
\langle \phi \rangle_{1,2}=\pm \sqrt{a+\sigma^2\,c_0}\,.
\end{equation}
This second set of solutions can only exist for $a>-\sigma^2\,c_0$.
Hence, the critical point is given in this case by
\begin{equation}
\label{eq:critpoint}
a_c=-\sigma^2\,c_0\,,
\end{equation}
in such a way that the ordered state appears for $a>a_c$.
The shift of the critical point increases the ordered region, due
to the effective multiplicative noise intensity $\sigma^2\,c_0$. 
This shift, as seen in the previous Sections, increases with increasing 
noise intensity and decreases for increasing correlation lengths. This
result coincides with the one given by a linear stability analysis of the
homogeneous state \cite{nises99,jordi98,becker94}, as expected. However, 
in contrast with the linear stability analysis, this calculation 
can be extended to
other situations and models not necessarily controlled by the linear term.

\subsection{Model B}

In this case, the stationary probability distribution (\ref{eq:pcolbg})
given by the mean--field approach for each phase and for  $D\to \infty$
is also  (\ref{eq:pst}) 
as can be seen using the steepest--descent method, as before. 
Following the procedure described above for model A, we formally integrate
now Eq.~(\ref{eq:pint}) to obtain an equation for $h$,
\begin{eqnarray}
\label{eq:fph}
h=-\langle f(\phi) \rangle_{\rm st} +2d\,\sigma^2(c_1
\,g(\langle \phi \rangle_{\rm st})\,\langle g'(\phi)\rangle_{\rm st}\,- 
\nonumber\\
\nonumber\\
-c_0\,\langle g(\phi)\, g'(\phi)\rangle_{\rm st})\,.
\end{eqnarray}   
At the limit $D\rightarrow\infty$, the averages appearing
in the previous expression are calculated using the stationary probability
distribution obtained above, leading to
\begin{equation}
\label{eq:bmedio4}
h=-f(\langle \phi \rangle_{\rm st})+ 
2d\,\sigma^2\,(c_1-c_0)\, g'(\langle \phi 
\rangle_{\rm st})\, g(\langle \phi \rangle_{\rm st})\,.
\end{equation} 

In the case $ a<a_T$, the field is homogeneous and we can replace in the 
above expression $\langle \phi \rangle_{\rm st} = \phi_0$.
Thus this equation gives us the value of $h$ in this case as a function 
of the initial condition. The results for model B and in the case defined by
(\ref{eq:fe})--(\ref{eq:ccol}) are plotted in Fig.~\ref{fig6} versus the
control parameter $a$, along with the values of $h$ given by mean--field
theory for finite but large $D$, obtained numerically in the previous
Section. We can see that these mean--field results approach
Eq.~(\ref{eq:bmedio4}) as $D$ increases, as it should be. The shift of
the transition point increases for increasing coupling strength as can be
seen from Fig.~\ref{fig6}.  

We now turn to the case $a>a_T$ where $h=0$. Now Eq.~(\ref{eq:bmedio4}) 
can be solved for $\langle \phi \rangle_{\rm st}$, which gives the 
values of the two bulk phases, 
\begin{equation}
\langle\phi\rangle_{1,2} = \pm \sqrt{ a + 2d\,\sigma^2\, (c_0-c_1)}\,,
\end{equation}
again for the particular model (\ref{eq:fe})--(\ref{eq:ccol}).
The transition line is determined by setting $\langle\phi\rangle_{1}=\phi_0$ in the
previous expression, which leads to 
\begin{equation}
\label{eq:at} 
a_T=\phi_0^2-2d\,\sigma^2(c_0-c_1)\,,
\end{equation}
and the critical point (for $\phi_0=0$) is then
\begin{equation}
\label{eq:acb}
a_c=-2d\,\sigma^2(c_0-c_1)\,.
\end{equation}
This result coincides with that coming from linear stability analysis
\cite{jordi98}. As in model A and in the previous Sections, the shift
is in the direction of increasing
the ordered region. Due to the factor $2d$ this shift is
larger than the one produced in model A for the same noise intensities.
Contrary to model A in the colored case, the shift does not depend only
on the effective multiplicative noise intensity $\sigma^2\,c_0$
but also on the noise correlation between first neighbors $c_1$,
which indicates the nontrivial influence of the spatial correlation
of the noise on conserved dynamics, as opposed to nonconserved dynamics
where this influence disappears in the mean--field approach.

\section{Conclusions}

Mean--field theory has been previously applied to nonconserved models
with additive and multiplicative white noises \cite{VPT,jordi96,broeck94}.
Here we have applied it in the case of spatially correlated 
multiplicative noise. Our mean--field results and numerical simulations of
the complete model in two dimensions indicate the decrease of the ordering 
role of multiplicative noise when its correlation length 
increases. 

We have also extended mean--field theory to deal with conserved models
by using the concept of a constant effective field.
As in the case of nonconserved systems, we have found that
additive noise has a disordering role, whereas multiplicative noise has an 
ordering one. The latter increases for increasing multiplicative noise 
intensity 
and for decreasing noise correlation length. However, the quantitative effects
of multiplicative noise are different in each model,  
the transition to order occurs earlier for model B than for model A. 
Moreover, mean field calculations show that
the correlation length of multiplicative noise has 
nontrivial effects in the conserved case, while for model A it just
decreases the effective noise intensity. Numerical simulations of the complete 
conserved model in two dimensions are in good agreement with mean--field 
predictions.

Previous works on model A with additive and multiplicative white noises 
have shown the presence of NIOTs and NIDTs. We have seen that, at least in the 
mean--field approach, these 
transitions appear for higher values of the noise intensity when 
multiplicative noise is spatially correlated. This is explained by the fact that
the effective noise intensity decreases.
Model B has also been found to go firstly through a NIOT and after through a
NIDT when the 
multiplicative noise intensity is increased. As in model A, the NIOT is
retarded when the correlation length of multiplicative noise increases. 
However, contrary to what happens in model A, the NIDT is advanced what
shows clearly different effects of noise correlation length upon conserved
and nonconserved models. 

Finally, in the strong--coupling limit we have found analytical
expressions for the critical--point shift and the steady--state bulk order parameter 
for both models A and
B with additive white and multiplicative colored noises. These
results coincide with previously reported predictions coming from
linear stability analysis \cite{jordi98}. 

\section{Acknowledgements}
This work has been supported by the Direcci\'on General de Investigaci\'on
Cient\'{\i}fica y T\'ecnica (Spain), under projects PB97--0141-C02-01, 
PB94--1167 and
PB96--0241. J.G.O. also acknowledges the Alexander von Humboldt--Stiftung
(Germany) for financial support, and thanks the group of Prof. Lutz
Schimansky--Geier in the Humboldt Universit\"at zu Berlin for their
hospitality.



\end{multicols}


{\large FIGURE CAPTIONS}

\begin{figure}
\caption{
\em Steady--state order parameter $\langle\phi\rangle_{\rm st}$ versus control
parameter $a$ for model A.
Lines are mean--field results and points correspond to numerical
simulations of the complete model for a two--dimensional square lattice
with $64\times 64$ cells, mesh size $\Delta x=1$,  
for system parameters $\lambda=0.0$ (circles
and solid line),
$\lambda=0.5$ (squares and dotted line) 
and $\lambda=1.5$ (triangles and dashed line). Other parameters
are $D=3.7$, $\varepsilon=0.1$ and $\sigma^2=5.0$.  
All simulations in this paper use the same lattice parameters. 
}
\label{fig1}
\end{figure}

\begin{figure}  
\caption{
\em Steady--state bulk order parameter $\langle\phi\rangle$ versus control
parameter $a$ for model B.
Lines are mean--field results and points correspond to numerical
simulations 
for $\lambda=0.0$ (circles and solid line),
$\lambda=0.5$ (squares and dotted line) 
and $\lambda=1.5$ (triangles and dashed line). Other parameters
are $D=3.7$, $\varepsilon=0.1$ and $\sigma^2=1.25$.}
\label{fig2}
\end{figure}

\begin{figure}
\caption{
\em 
Constant effective field $h$ as a function of the control parameter $a$, 
as obtained from mean--field theory, for $D=3.7$ and $\phi_0=0.2$. All the
cases with multiplicative noise have also additive noise with $\epsilon=0.1$.} 
\label{fig3}
\end{figure}

\begin{figure}  
\caption{
\em Steady--state bulk order parameters 
$\langle\phi\rangle$ versus control
parameter $a$ for model A and B with additive white and multiplicative colored noises
for different correlation lengths.
Lines are mean--field results for model A with $\lambda=0.0$ (solid line) and
$\lambda=1.5$ (dotted line) and
for model B with $\lambda=0.0$ (solid line) 
and $\lambda=1.5$ (dashed line). Points correspond to numerical
simulations of model A (empty symbols) and model B (full symbols).
Of these, circles correspond
to white multiplicative noise and triangles to $\lambda=1.5$.
Other parameters
are $D=3.7$, $\varepsilon=0.1$, 
$\sigma^2_A=5$ and $\sigma^2_B=1.25$.
}
\label{fig4}
\end{figure}

\begin{figure}
\caption{
\em 
Mean field steady--state bulk order parameters $\langle\phi\rangle$
versus multiplicative noise intensity for models A with $\lambda=0.0$ (solid line)
and $\lambda=0.5$ (dotted line) and for model B with $\lambda=0.0$ (solid 
line) and $\lambda=0.5$ (dashed line).
Parameters are $\epsilon=1$, $a=0.75$ and $D=2.66$.}
\label{fig5}
\end{figure}

\begin{figure}
\caption{
\em Constant effective field $h$ versus control parameter $a$ for
$D\to\infty$ (solid line) as given by Eq.~(\protect\ref{eq:bmedio4}),
and for $D=3.7$ (dashed line) and $D=20$ (dotted line) coming from
the mean--field approach described in Section IV.B. Other parameters
are $\phi_0=0.2$, $\epsilon=0.1$, $\sigma^2=1.25$ and $\lambda=0.5$.}
\label{fig6}
\end{figure}

\newpage

\centerline{\psfig{figure=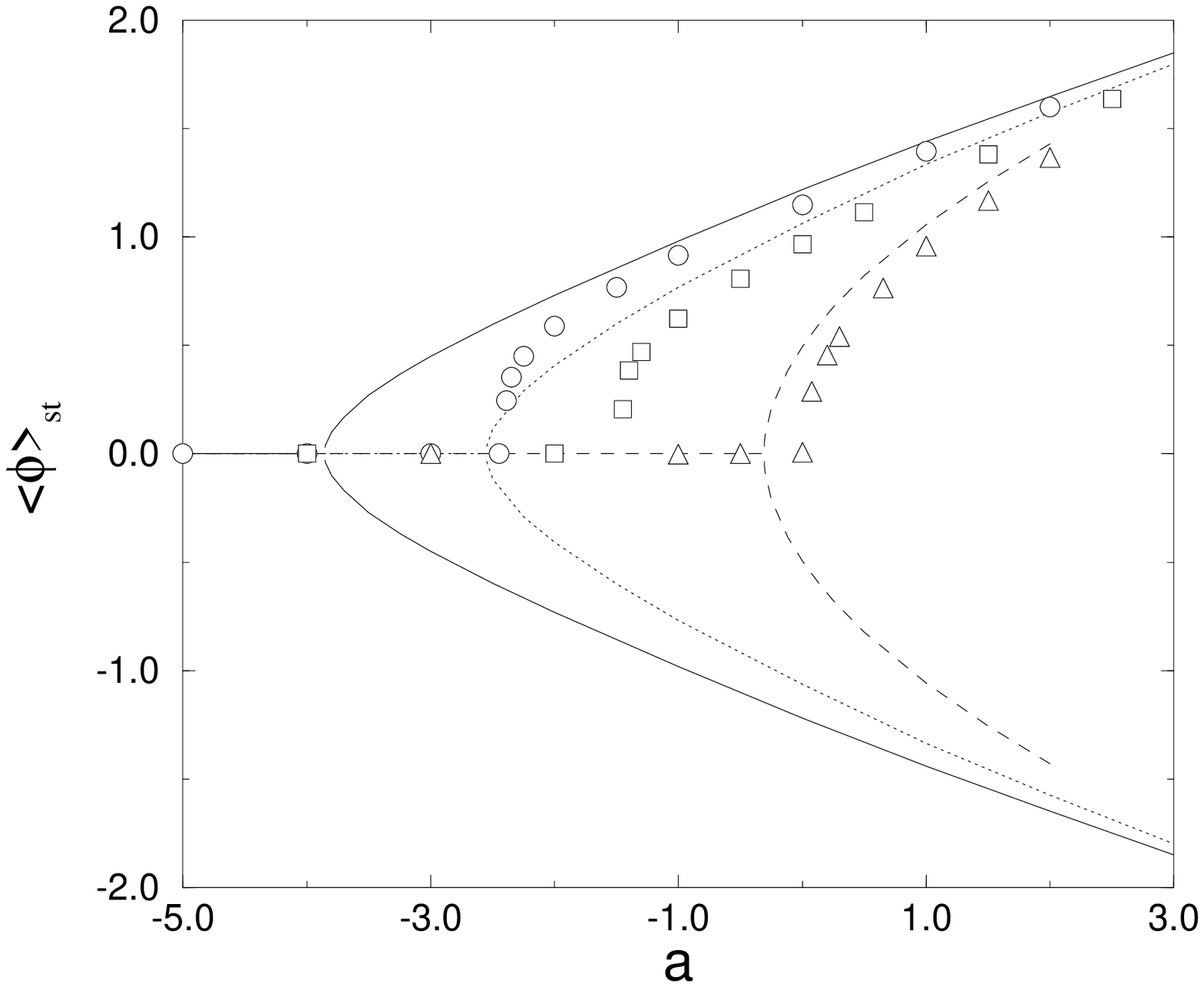,width=9.5cm,height=7cm}}
\centerline{FIGURE 1}
\centerline{Noise--Induced Phase Separation: Mean Field Results}
\centerline{M. Iba\~nes et al.}

\centerline{\psfig{figure=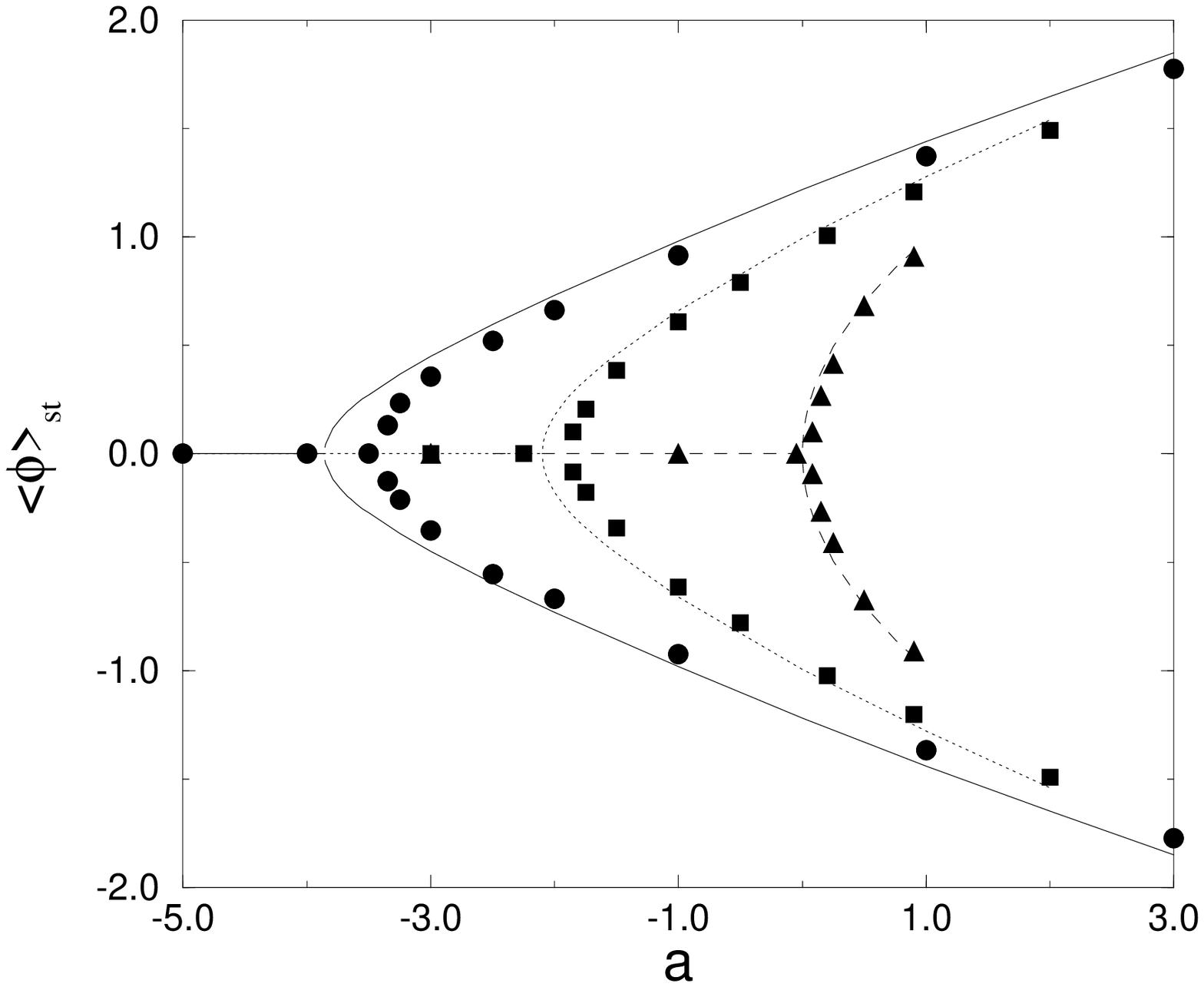,width=9.5cm,height=7cm}}
\centerline{FIGURE 2}
\centerline{Noise--Induced Phase Separation: Mean Field Results}
\centerline{M. Iba\~nes et al.}

\centerline{\psfig{figure=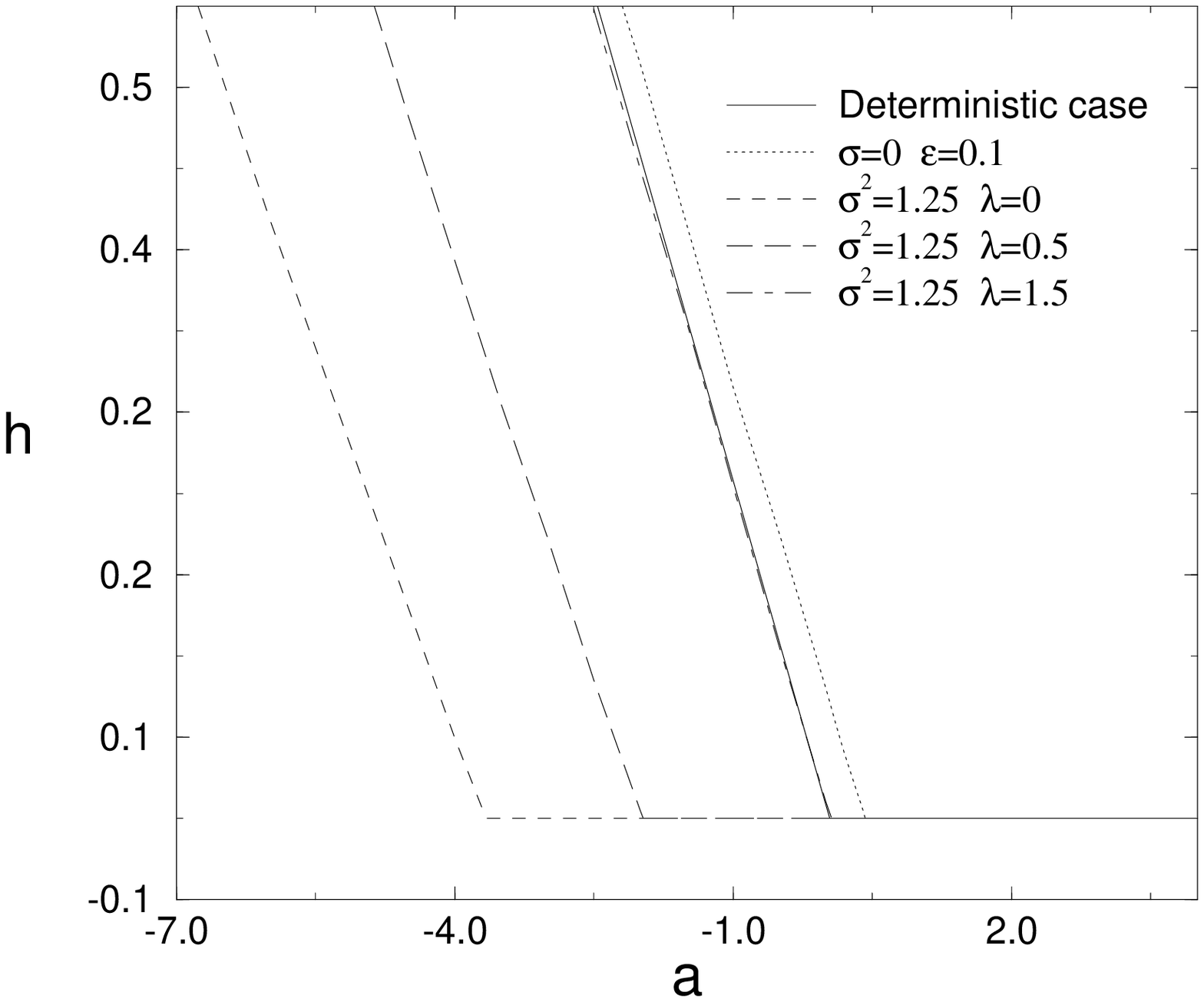,width=9.5cm,height=7cm}}
\centerline{FIGURE 3}
\centerline{Noise--Induced Phase Separation: Mean Field Results}
\centerline{M. Iba\~nes et al.}

\centerline{\psfig{figure=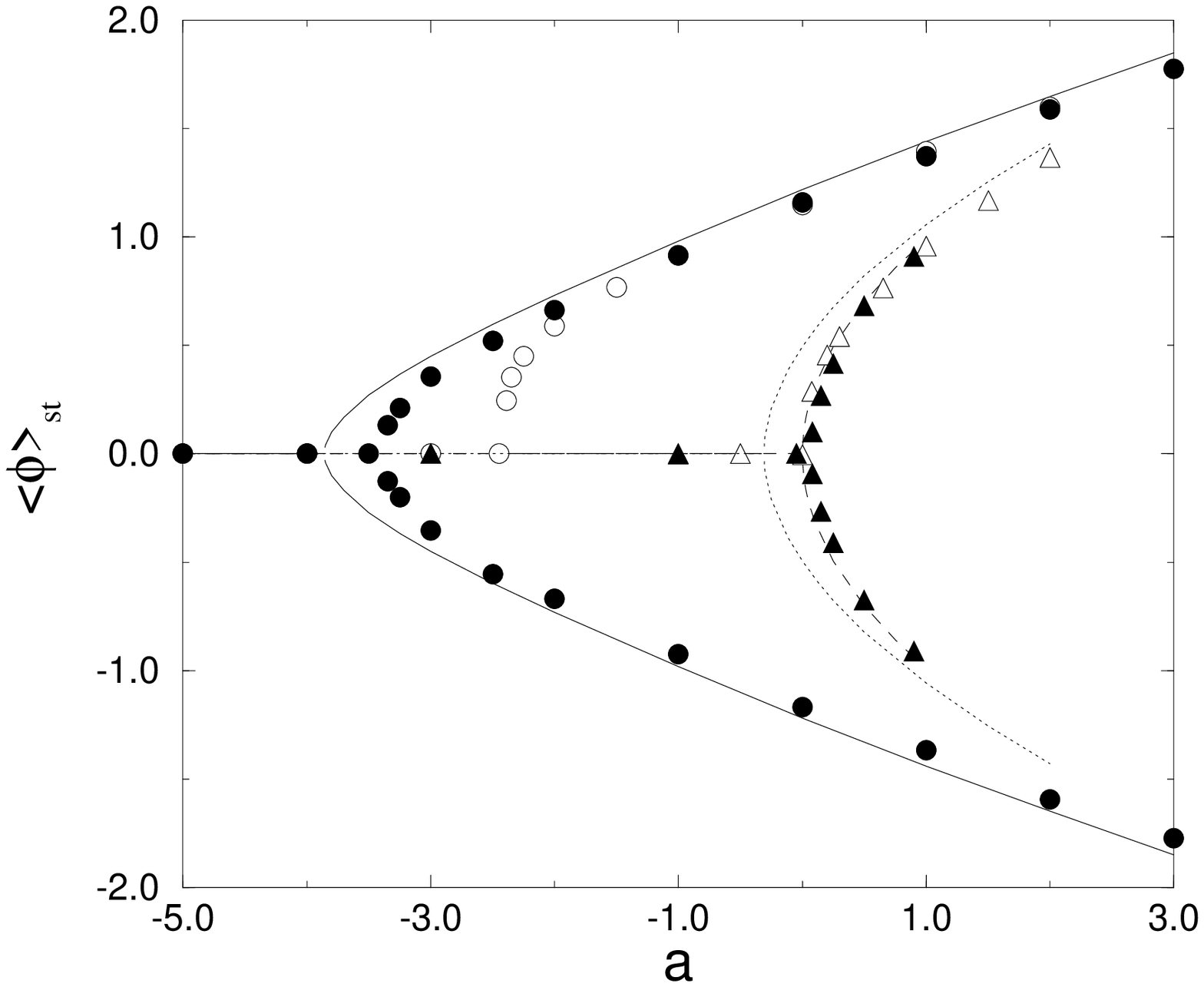,width=9.5cm,height=7cm}}
\centerline{FIGURE 4}
\centerline{Noise--Induced Phase Separation: Mean Field Results}
\centerline{M. Iba\~nes et al.}

\newpage
\centerline{\psfig{figure=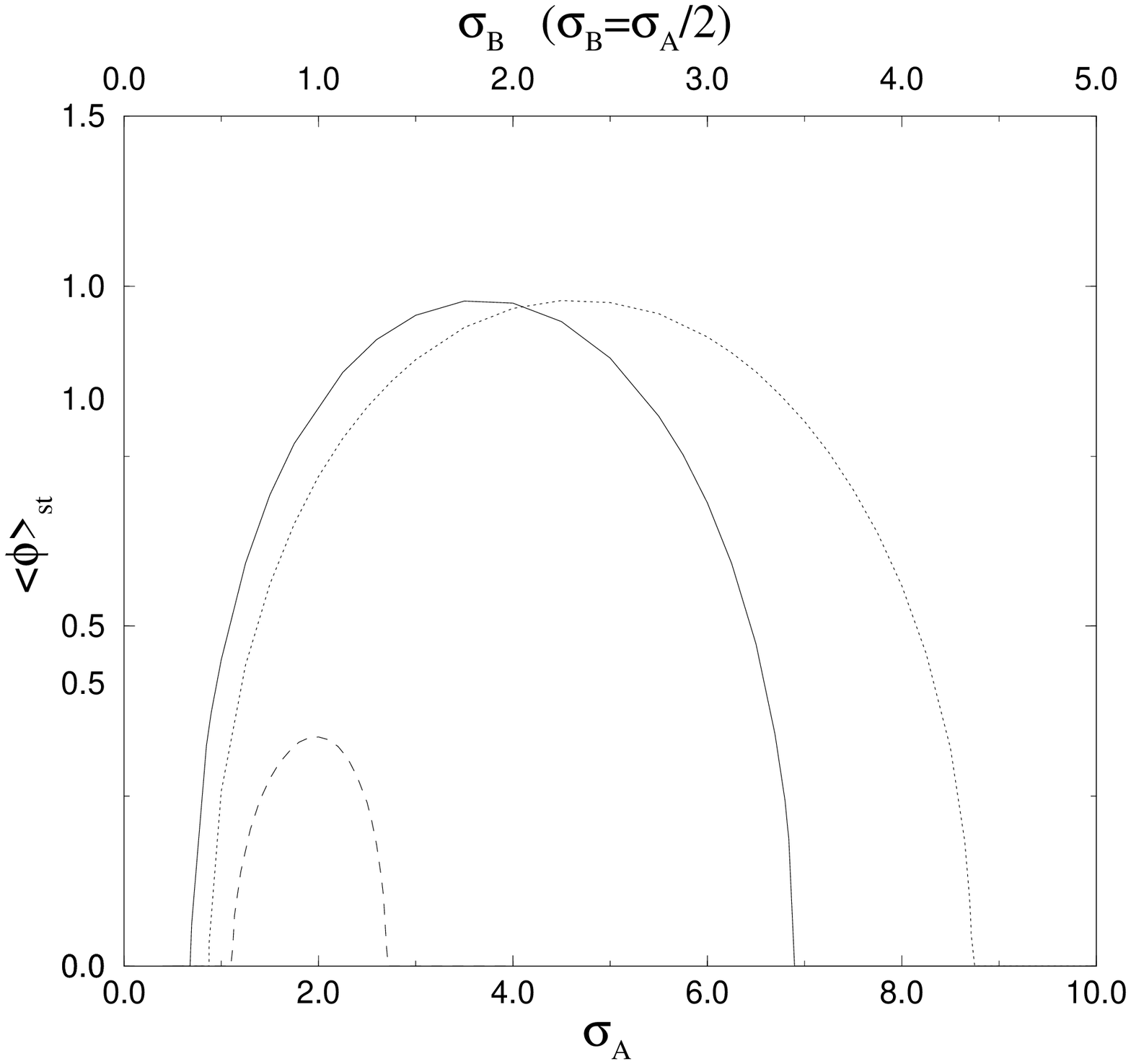,width=9cm,height=6cm}}
\centerline{FIGURE 5}
\centerline{Noise--Induced Phase Separation: Mean Field Results}
\centerline{M. Iba\~nes et al.}

\centerline{\psfig{figure=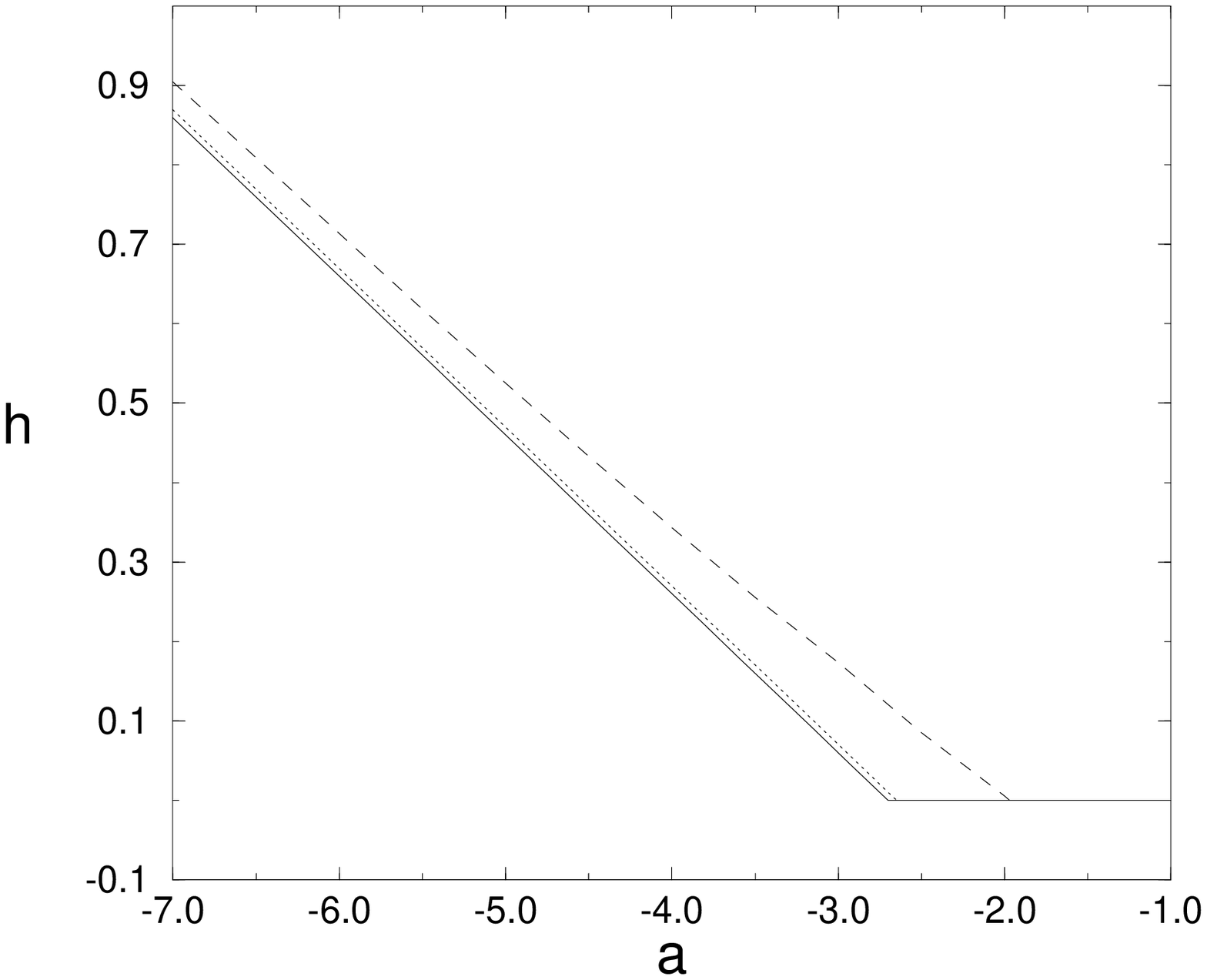,width=9.5cm,height=7cm}}
\centerline{FIGURE 6}
\centerline{Noise--Induced Phase Separation: Mean Field Results}
\centerline{M. Iba\~nes et al.}


\begin{references}
\bibitem{horsthemke84} W. Horsthemke, R. Lefever, {\it Noise-Induced 
Transitions} (Springer, Berlin, 1984).
\bibitem{gamma98} L.~Gammaitoni, P.~H\"anggi, P.~Jung, and F.~Marchesoni,
Rev. Mod. Phys. {\bf 70}, 223 (1998).
\bibitem{hanggi96} P.~H\"anggi and R.~Bartussek,
in {\em Nonlinear Physics of Complex Systems}, ed. by J.~Parisi,
S.C. M\"uller, and W.~Zimmermann (Springer, Berlin, 1996).
\bibitem{nises99} J.~Garc\'{\i}a--Ojalvo and J.M.~Sancho,
{\em Noise in Spatially Extended Systems}, to appear
(Springer, New York, 1999).
\bibitem{ojalvo93} J. Garc{\'\i}a-Ojalvo, A. Hern\'andez--Machado
and J.M. Sancho, Phys. Rev. Lett. {\bf 71}, 1542 (1993).
\bibitem{parr96} J.M.R. Parrondo, C. Van den Broeck, J. Buceta and
F.J. de la Rubia, Physica {\bf A 224}, 153 (1996).
\bibitem{VPT} C. Van den Broeck, J.M.R. Parrondo and R. Toral,
Phys. Rev. Lett. {\bf 73}, 3395 (1994); C. Van den Broeck, J.M.R.
Parrondo, R. Toral and R. Kawai, Phys. Rev. E {\bf 55}, 4084 (1997).
\bibitem{jordi96} J. Garc{\'\i}a-Ojalvo, J.M.R. Parrondo, J.M. Sancho
and C. Van den Broeck, Phys. Rev. E {\bf 54}, 6918 (1996).
\bibitem{man97} S. Mangioni, R. Deza, H. Wio, R. Toral. Phys. Rev. Lett. {\bf 79}, 2389 
(1997).
\bibitem{SSR1} P.~Jung and G.~Mayer--Kress, Phys.~Rev.~Lett. {\bf 74},
2134 (1995).
\bibitem{SSR2} F. Marchesoni, L. Gammaitoni and A.R. Bulsara, Phys. Rev.
Lett. {\bf 76}, 2609 (1996).
\bibitem{santos99} M.A. Santos and J.M. Sancho, Phys. Rev. E {\bf 59}, 98
(1999). 
\bibitem{exc} S. K\'ad\'ar, J. Wang and K. Showalter, Nature {\bf 391},
770 (1998); P. Jung, A. Cornell--Bell, F. Moss, S. K\'ad\'ar, J. Wang
and K. Showalter, Chaos {\bf 8}, 567 (1998); J. Wang, S. K\'ad\'ar, P. Jung
and K. Showalter, Phys. Rev. Lett. {\bf 82}, 855 (1999). 
\bibitem{dei89} R.J.Deissler, J. Stat. Phys., {\bf 54}, 1459 (1989).
\bibitem{SCSW} M. Santagiustina, P. Colet, M. San Miguel and D. Walgraef,
Phys. Rev. Lett. {\bf 79}, 3633 (1997); {\em ibid.}, Phys. Rev. E. {\bf 58},
3843 (1998).
\bibitem{jordi98} J. Garc\'{\i}a--Ojalvo, A.M. Lacasta, J.M. Sancho and
R. Toral, Europhys. Lett. {\bf 42}, 125 (1998).
\bibitem{becker94}  A. Becker and L. Kramer,  Phys. Rev. Lett.
{\bf 73}, 955 (1994); {\em ibid.}, Physica D {\bf 90}, 408
(1995).
\bibitem{DRG1} G. Grinstein, M.A. Mu\~noz and Y. Tu, Phys. Rev. Lett.
{\bf 76}, 4376 (1996).
\bibitem{DRG2} J.M. Sancho, J. Garc\'{\i}a-Ojalvo and H. Guo,
Physica D {\bf 113}, 331 (1998).
\bibitem{MNUC1} Y. Tu, G. Grinstein and M.A. Mu\~noz, Phys. Rev. Lett.
{\bf 78}, 274 (1997).
\bibitem{MNUC2} W. Genovese, M.A. Mu\~noz and J.M. Sancho,
Phys. Rev. E {\bf 57}, R2495 (1998).
\bibitem{broeck94} C. Van den Broeck, J.M.R. Parrondo, J. Armero and 
A. Hern\'andez-Machado, Phys. Rev. E {\bf 49}, 2639 (1994).
\bibitem{zaikin} P.S. Landa, A.A. Zaikin and L. Schimansky--Geier,
Chaos, Solitons and Fractals {\bf 9}, 1367 (1998); A.A. Zaikin and
L. Schimansky--Geier, Phys. Rev. E {\bf 58}, 4355 (1998).
\bibitem{Ito} Throughout this paper we follow the Stratonovich interpretation.
In the Ito interpretation, where the homogeneous state is linearly stable,
the noise--induced transitions reported in this paper will not appear.
\bibitem{gunton83} J.D. Gunton, M. San Miguel and P.S. Sahni,
in {\em Phase Transitions and Critical Phenomena}, Vol. 8, edited
by C. Domb and J.L. Lebowitz (Academic Press, New York, 1983).
\bibitem{text} In this assumption local correlations are discarded.
A second--order correction can be implemented as described in 
\cite{VPT}.
\bibitem{miguela} M.A. Mu\~noz, U. Marini and R. Cafiero,
Europhys. Lett. {\bf 43}, 552 (1998).

\end{references}
\end{document}